\documentclass[aps,pra,reprint,showpacs,floatfix,twocolumn,nofootinbib]{revtex4-1}
\usepackage{amsmath,amsthm,amsfonts,amssymb,amscd}
\usepackage{t1enc}
\usepackage[utf8]{inputenc}
\usepackage[english]{babel}
\usepackage{babel}
\usepackage{graphicx}
\usepackage[T1]{fontenc}
\usepackage{latexsym}
\usepackage{enumerate}
\usepackage[caption=false]{subfig}

\begin{document}

\title{Fine-tuning of orbital angular momentum in an optical parametric oscillator} 
\author{R. F. Barros, G. B. Alves, D. S. Tasca, C. E. R. Souza, A. Z. Khoury}
\address{Instituto de F\'isica, Universidade Federal Fluminense, CEP 24210-346, Niter\'oi-RJ, Brazil}
\email{rafael.fprb@gmail.com}
\begin{abstract}
We investigate the dynamical properties of a type-II optical parametric oscillator pumped by a structured light beam carrying orbital angular momentum. Different dynamical regimes are theoretically derived and experimentally demonstrated, depending on the anisotropy effects imposed by the nonlinear medium. In a weakly anisotropic regime, fine-tuning of the orbital angular momentum transfer is obtained by adjusting the crystal parameters. Under strong anisotropy, orbital angular momentum transfer is not possible and a sharp dynamical behavior is observed, with abrupt switching between different transverse modes.
\end{abstract}

\maketitle

\section{Introduction}\label{sec-introduction}
 
The transverse structure of light is an important degree of freedom in many optical systems. In particular, the topic of vortex beams and the orbital angular momentum (OAM) of light has attracted considerable interest in the recent years, leading to a growing number of devices and applications, such as optical vortex micromachining \cite{PhysRevLett.110.143603,omatsu:nano}, quantum teleportation and quantum criptography\cite{PhysRevA.77.032345,dambrosio_complete_2012}. Some authors have observed complex selection rules in the nonlinear wave mixing with structured light beams, as in second-harmonic generation \cite{PhysRevA.96.053856} and parametric amplification \cite{Fang:19}, including but not limited to the OAM conservation. 

The role of transverse mode dynamics has also been studied in optical parametric oscillators (OPO) \cite{Aadhi:17,PhysRevA.70.013812}. OPOs are well-known sources of non-classical states of light \cite{PhysRevLett.59.2555}, such as squeezed and entangled states \cite{Mertz:91,PhysRevLett.68.3663,Keller:08,Bernardo2005}. Quantum information protocols based on continuous variables have been proposed in \cite{braunstein2012quantum} and experimentally realized in \cite{PhysRevLett.112.120505,PhysRevLett.107.030505} by the generation of entangled frequency combs in an OPO. Furthermore, the parametric oscillation also presents a highly nonlinear dynamics, in analogy to other physical, biological and social systems. It has been predicted theoretically that a triply resonant OPO may present Hopf bifurcations, chaos and self-pulsing behavior \cite{Lugiato1988}. Fast oscillations and spontaneous symmetry breaking in such systems were observed in \cite{SURET:2001} and \cite{Longchambon:05}, respectively.

One key feature of the nonlinear wave mixing in an OPO is that the interplay between gain and losses is dictated by cavity detunings \cite{debuisschert_type-ii_1993}. This is particularly important in the case of structured light beams, once different transverse modes are spectrally separated due to the Gouy phase. We have shown in a recent work that the mode detunings may overcome the transverse coupling in the definition of the threshold hierarchy \cite{PhysRevA.98.063825}. Nevertheless, the transverse mode dynamics in an OPO is drastically affected by anisotropy. In \cite{PhysRevA.70.013812}, it was shown that astigmatism may prevent OAM transfer to one of the down-converted beams. This could only be overcome in OPOs containing periodically poled \cite{Sharma:18,Aadhi:17} or type-I \cite{Abulikemu:15} crystals.

In this paper,  we demonstrate that the astigmatism introduced by the crystal birefringence can be suitably controlled to produce two different operating regimes in the OPO. One can either fine-tune the OAM transfer to the down-converted beam with extraordinary polarization or induce an abrupt Hermite-Gaussian mode-switching in the ordinary polarization. These results can be useful for building spatially structured beams for classical and quantum applications. 

\section{Experimental setup}\label{experimental}

\begin{figure}[b]
	\includegraphics[scale=0.36]{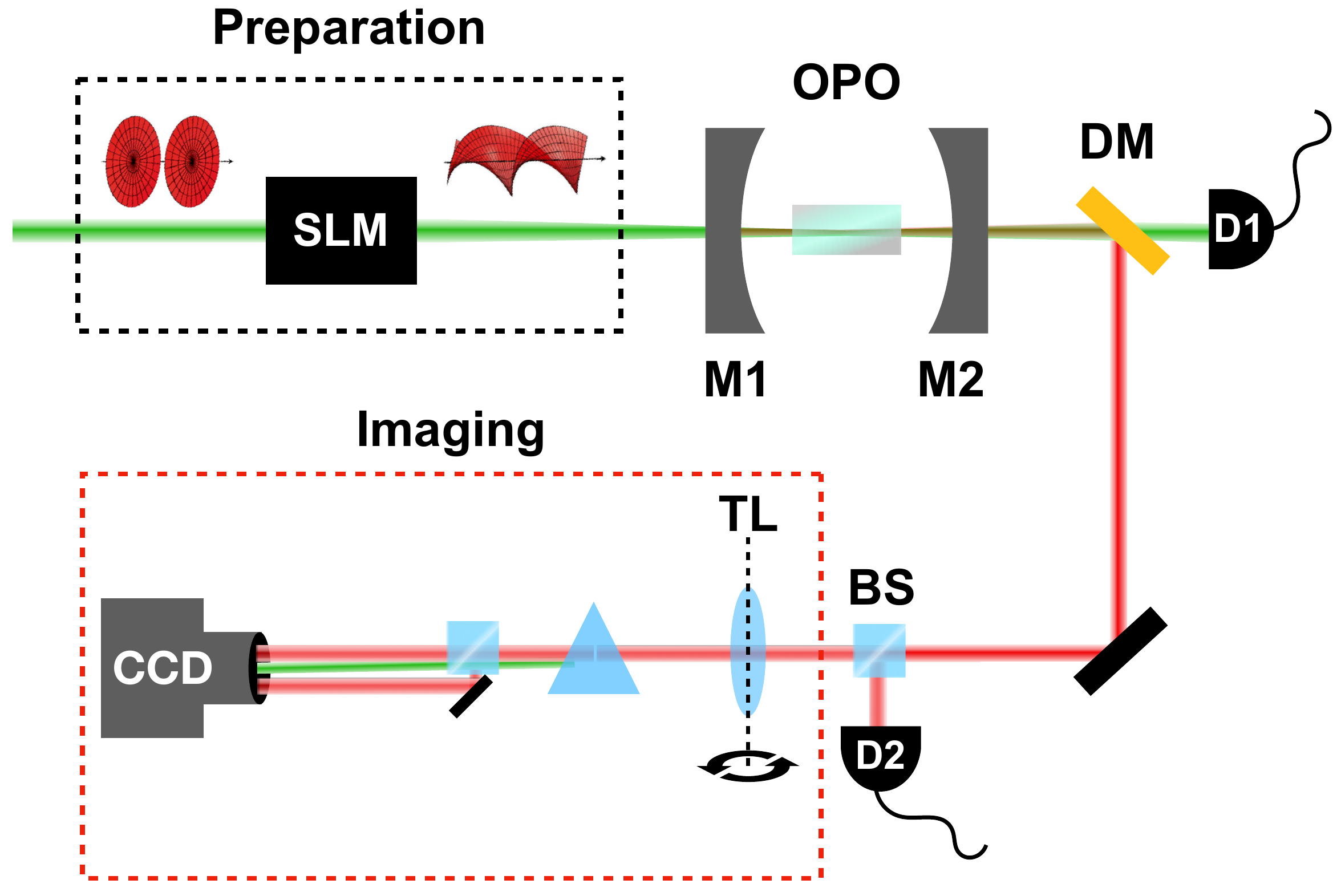} 
	\caption{Sketch of the experimental setup. SLM, spatial light modulator; M1 and M2, cavity mirrors; DM, dichroic mirror; D1 and D2, detectors; TL, tilting lens; CCD, charge-coupled device.}
	\label{setup}
\end{figure}

The experimental setup is sketched in figure \ref{setup}. The OPO is formed by two concave mirrors M1 (HR@1064nm, 96\%@532nm, R=-25mm) and M2 (HR@1064nm+532nm, R=-25mm) and a 5mm  long type-II KTP crystal (by Altechna) in a linear and nearly confocal configuration. The crystal is temperature-controlled by a Peltier element and mirror M2 is mounted over a piezoelectric ceramic (PZT) for precision control of the cavity length. The pump beam is a frequency-doubled TEM$_{00}$ Nd:Yag laser (InnoLight GmbH laser, Diabolo product line) which is converted to a LG$_{01}$ ($p=0$, $l=1$) by a spatial light modulator (SLM - Hamamatsu, model X10468-01).

The OPO output is filtered by a dichroic mirror (DM) which reflects most of the 532nm intensity to a detector (Thorlabs, DET100A) for monitoring of the pump resonances. The remaining pump and the converted beams are focused by TL and taken in a charge-coupled device (CCD) camera. The lens can also be tilted \cite{VAITY20131154} to allow topological charge measurements with the CCD. Part of the IR is separated in a 50/50 non-polarizing beam-splitter (BS1) and detected at D2.

By measuring the cavity's finesse one can extract the decay rates $\gamma\approx 5\,mrad$ and $\gamma_p\approx 90\,mrad$ at 532\,nm and 1064\,nm, respectively, and estimate the accumulated round trip phase difference between the Hermite-Gaussian ($HG_{mn}$) components of the signal (ordinary-$o$) and idler (extraordinary-$e$) beams
\begin{eqnarray}
|\varphi^o_{10}-\varphi^o_{01}| &\approx & 4.3 \gamma \;,\label{eq:split1}\\
|\varphi^e_{10}-\varphi^e_{01}| &\approx & 0.6 \gamma \;. \label{eq:split2}
\end{eqnarray}
Notice that the astigmatism is much more pronounced for the $o$-polarized wave. 

In order to investigate the multimode operation of the OPO under a LG$_{01}$ pump, we take advantage of the dependence of the refractive indexes on the temperature and angle of incidence to produce a fine-tuning of the crystal's anisotropy. This procedure changes the astigmatic mode splitting and allows us to access different operating regimes of the parametric oscillation.

Our investigation will be separated in two situations. In the first, the signal beam (weak astigmatism) is excited in the $\{HG_{10},HG_{01}\}$ subspace and couples with a TEM$_{00}$ idler. In the second, the idler beam (strong astigmatism) is excited in the $\{HG_{10},HG_{01}\}$ subspace and couples with a TEM$_{00}$ signal. These two cases entail totally different dynamics in what regards OAM transfer. While the first operating condition allows fine tuning of the signal OAM, the second one exhibits a sharp behavior with abrupt switching between $HG_{10}$ and $HG_{01}$ operation in the idler, preventing any type of superposition between the two modes. As we will see, the important figure of merit determining the two behaviors is the relative magnitude of the astigmatic mode splitting in terms of the cavity linewidth as expressed by (\ref{eq:split1}) and (\ref{eq:split2}). This will naturally come out from the theoretical model discussed in the next section.

\section{Dynamical equations and steady state} \label{anisotropy}

The OPO equations entail a complex dynamical structure involving a network of transverse modes coupled by the three-mode overlap integrals \cite{Schwob1998}. However, the threshold hierarchy imposed by the spatial overlap and boundary conditions limits the number of effectively operating modes. When written in the LG basis, for example, it can be shown that these integrals ensure OAM conservation in the downconversion process. Experimental realizations of complete OAM transfer have already been reported for OPOs containing periodically poled crystals \cite{Aadhi:17}.

This OAM exchange, however, can be dramatically affected by the crystal anisotropy in type-II OPOs. In these media, the different higher-order components in the transverse structure of a given polarization experience distinct cavity lengths, causing a rotational symmetry breaking and preventing the OAM exchange to occur in most cases. Moreover, this symmetry breaking depends drastically on the polarization, being much more pronounced in one of the two down-converted beams. For this reason, in \cite{PhysRevA.70.013812} OAM transfer could only be achieved from the pump to the down-converted beam carrying the same polarization as the pump. 

In order to study the multimode parametric oscillation in anisotropic media, let us consider the case where the pump is structured with an arbitrary superposition of 1st order HG modes, 
while one of the down-converted beams oscillates in a TEM$_{00}$ mode and the other is a superposition of 1st order HG modes. 
For the moment, we will not specify the polarizations associated with each transverse mode and leave this assignment to the discussion about the polarization-induced astigmatism.
The structured pump beam is represented by 
\begin{equation}
E_{in}=\sqrt{I_{in}}(\cos\frac{\theta}{2} HG_{10}+e^{i\phi}\sin\frac{\theta}{2} HG_{01}) \;,
\label{pump}
\end{equation}
where $0\leq \theta\leq \pi$ and $0\leq \phi \leq 2\pi\,$. 
In this case, the OPO dynamics is described by the following dynamical equations for the complex amplitudes
\begin{align}
&\frac{d\alpha_{ph}}{dt} = -\gamma_p\left(1-i\Delta_{ph}\right)\alpha_{ph} + ig\alpha_h\alpha_{00} +t_p\cos\frac{\theta}{2} \sqrt{I_{in}} \label{eq-5OPO1} \\
&\frac{d\alpha_{pv}}{dt} = -\gamma_p\left(1-i\Delta_{pv}\right)\alpha_{pv} + ig\alpha_v\alpha_{00} +t_pe^{i\phi}\sin\frac{\theta}{2}\sqrt{I_{in}}\label{eq-5OPO2} \\
&\frac{d\alpha_h}{dt} = -\gamma\left(1-i\Delta_h\right)\alpha_h + ig\alpha_{ph}\alpha^{*}_{00}\label{eq-5OPO3} \\
&\frac{d\alpha_v}{dt} = -\gamma\left(1-i\Delta_v\right)\alpha_v + ig\alpha_{pv}\alpha^{*}_{00}\label{eq-5OPO4} \\
&\frac{d\alpha_{00}}{dt} = -\gamma\left(1-i\Delta_{00}\right)\alpha_{00} + ig\alpha_{ph}\alpha^{*}_h+ig\alpha_{pv}\alpha^{*}_v\label{eq-5OPO5}, 
\end{align}
where the sub-indexes $h$ and $v$ were assigned to the HG$_{10}$ and HG$_{01}$ complex amplitudes, respectively, and the sub-index $00$ was assigned to the TEM$_{00}$ mode amplitude. In these equations $g$ is the nonlinear coupling constant, 
$\Delta_j=\delta\varphi_j/\gamma_j$ ($j=h,v,00$) are the normalized detunings for the down-converted modes and $\Delta_{pj}=\delta\varphi_{pj}/\gamma_{pj}$ ($j=h,v$) are the normalized detunings for the pump. 

\begin{figure*}[t]
	\includegraphics[scale=0.25]{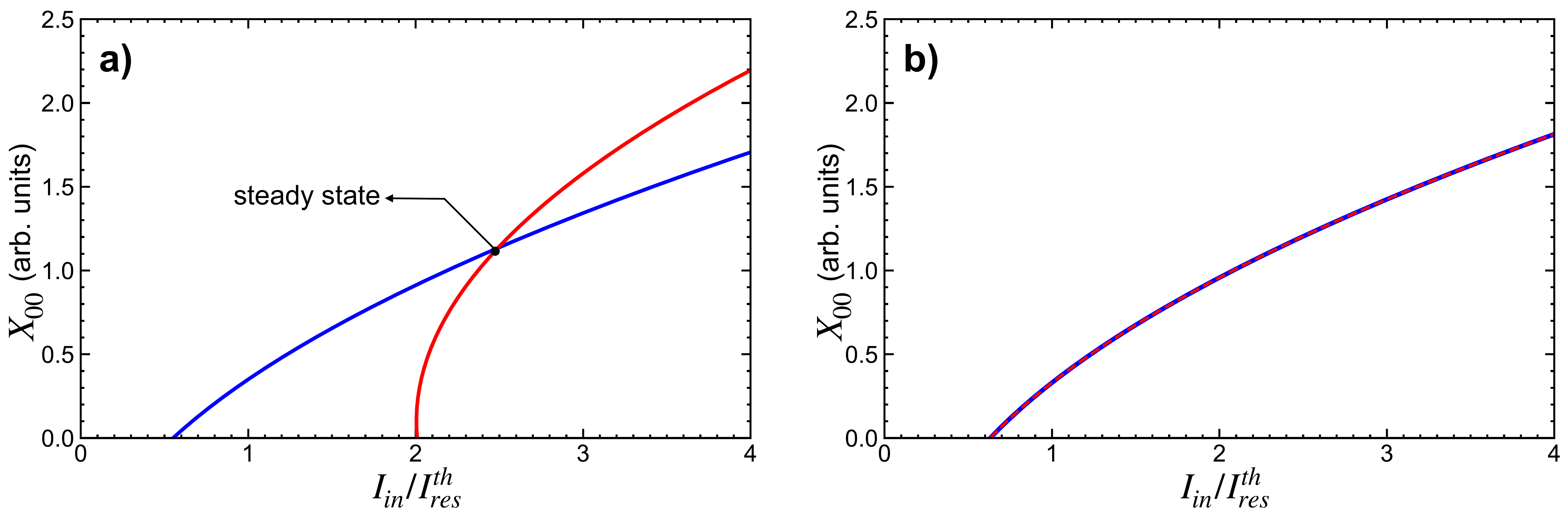}
	\caption{Graphical representation of the polynomial relations given by \eqref{idler-h} (blue) and \eqref{idler-v} (red) for two different sets of parameters. 
		(a) $\Delta_h=1$, $\Delta_v=-0.1$,  $\Delta_{ph}=1.1$,  $\Delta_{pv}=0$,  $\Delta_{00}=0.4\,$. The steady state solution is given by the intersection between the two parabolas.
		(b) $\Delta_h=-\Delta_v=0.5$, $\Delta_{ph}=-\Delta_{pv}=0.1$, $\Delta_{00}=0\,$. In this case, conditions \eqref{produto} and \eqref{modulo} are fulfilled, and  \eqref{idler-h} and \eqref{idler-v} define a single parabola determining the steady state solution for any pump level.}
	\label{parabolas}
\end{figure*}

Steady-state solutions for \eqref{eq-5OPO1}-\eqref{eq-5OPO5} can only be obtained when the detunings of the down-converted beams are either all equal or all different from each other. The first case ($\Delta_{00}=\Delta_h=\Delta_v$), describes an isotropic system with no astigmatism, which does not correspond to the actual physical situation. In this particular case, equations \eqref{eq-5OPO1}-\eqref{eq-5OPO5} can be reduced to a set of three equations by a suitable redefinition of the complex amplitudes, where the generated 1st order mode simply mimics the pump input. 

The situation of physical interest is when all detunings are different, which better describes the astigmatic OPO.  In this case, substituting \eqref{eq-5OPO3} and \eqref{eq-5OPO4} in \eqref{eq-5OPO5} lead to the following solutions for the intracavity pump intensity
 \begin{eqnarray}
|\alpha_{ph}|^2=\frac{\gamma^2}{g^2} \Theta_h(1 + \Delta_h^2)\label{pump-h-intra} \\
|\alpha_{pv}|^2=\frac{\gamma^2}{g^2} \Theta_v(1 + \Delta_v^2)\label{pump-v-intra} ,
 \end{eqnarray}
where
 \begin{eqnarray}
 \Theta_h &=& \frac{\Delta_{00}-\Delta_v}{\Delta_h-\Delta_v} \;,
 \nonumber\\
 \Theta_v &=& \frac{\Delta_h-\Delta_{00}}{\Delta_h-\Delta_v}\label{theta} \;.
 \end{eqnarray}
Notice that $\Delta_{00}$ must lie in between $\Delta_h$ and $\Delta_v\,$ ($\Delta_{h(v)}\leq\Delta_{00}\leq\Delta_{v(h)}$), so the right-hand sides of \eqref{pump-h-intra} and \eqref{pump-v-intra} are positive, 
which is a major restriction over the detuning parameters. 
From \eqref{eq-5OPO3} and \eqref{eq-5OPO4}, the $h$ and $v$ down-converted intensities are simply
\begin{equation}
|\alpha_{j}|^2 = \Theta_j |\alpha_{00}|^2\qquad (j=h,v)\;.
\end{equation}

Now we substitute \eqref{pump-h-intra} and \eqref{pump-v-intra} back into \eqref{eq-5OPO1} and \eqref{eq-5OPO2}, obtaining two different equations relating the intensities of the TEM$_{00}$ and the pump input
\begin{eqnarray}
\frac{I_{in}}{I_{res}^{th}} &=& \frac{\Theta_h}{\cos^2{\theta}/{2}}\left(X_{00}^2+B_h X_{00}+C_h\right)\;,
\label{idler-h}\\
\frac{I_{in}}{I_{res}^{th}} &=& \frac{\Theta_v}{\sin^2{\theta}/{2}}\left(X_{00}^2+B_v X_{00}+C_v\right)\;,
\label{idler-v}
\end{eqnarray}
where
\begin{equation}
X_{00} \equiv \frac{g^2}{\gamma\gamma_p}|\alpha_{00}|^2\qquad \textrm{and}\qquad I_{res}^{th} \equiv \frac{\gamma^2\gamma_p^2}{g^2t_p^2}\;,
\end{equation}
and the coefficients $B_j$ and $C_j$ ($j=h,v$) are defined as
\begin{eqnarray}
B_j &=& 2(1-\Delta_{pj} \Delta_{j})\;,\label{b-coef}
\\
C_j &=& (1+\Delta_{pj}^2)(1+\Delta_{j}^2)\;.\label{c-coef}
\end{eqnarray}

Notice that \eqref{idler-h} and \eqref{idler-v} implicitly define two different polynomials for the TEM$_{00}$ intensity as a function of the pump input intensity $I_{in}$, whose parameters depend on the mode detunings. For a stationary solution to exist with a given pump level $I_{in}\,$, both equations must be satisfied by the same value of $X_{00}\,$. This results in a complicated condition for the set of parameters $\Delta_{pj}, \Delta_j,\Delta_{00}$ and  $\theta\,$, and an analytical expression is not straightforward. The stationary solution must be obtained numerically as the graphical intersection between the corresponding parabolas, as depicted in figure \ref{parabolas}a. 

A much simpler situation occurs when the polynomials in \eqref{idler-h} and \eqref{idler-v} are identical, thus defining a single parabola as shown in figure \ref{parabolas}b. 
In this case, a stationary solution exists for every pump value above threshold, given that the following restrictions are satisfied
\begin{eqnarray}
\Theta_v &=& \Theta_h \tan^2\theta/2,
\nonumber\\
B_h &=& B_v
\label{equal-thetas}\\
C_h &=& C_v\;.
\nonumber
\end{eqnarray}
The first restriction gives $\Delta_{00}$ as the average between $\Delta_{h}$ and $\Delta_{v}$ weighted by the relative intensities of the pump input,
\begin{equation}
\Delta_{00}={\Delta_h\cos^2\frac{\theta}{2}+\Delta_v\sin^2\frac{\theta}{2}}\;. \label{Delta-idler}
\end{equation}
%

\begin{figure*}
	\includegraphics[scale=0.28]{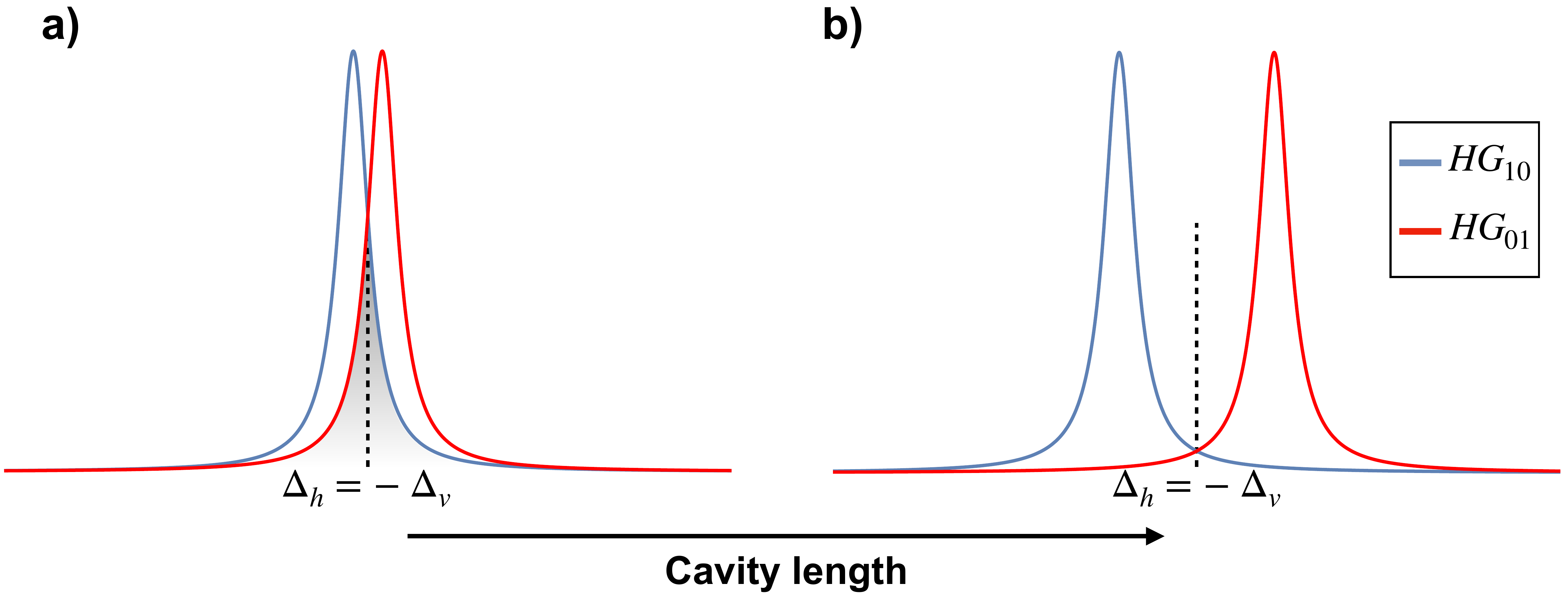}
	\caption{Illustration of the astigmatic mode splitting for a) signal ($\Delta=0.3\gamma$) and b) idler ($\Delta=2.1\gamma$) polarizations, according to \eqref{eq:split1} and \eqref{eq:split2}. The vertical dashed lines indicate the operating regime described by \eqref{typeii}.}
	\label{astigmatism}
\end{figure*}

The other two restrictions lead to
\begin{eqnarray}
\Delta_{ph}\Delta_h&=&\Delta_{pv}\Delta_v \;, \label{produto}\\ 
\Delta^2_{ph}+\Delta^{2}_h&=&\Delta^{2}_{pv}+\Delta^{2}_v \;.\label{modulo}
\end{eqnarray}
This system has four different solutions, which are described in the following. The first one is given by $\{\Delta_h=\Delta_v,\Delta_{ph}=\Delta_{pv}\}$, which is excluded by our initial hypothesis of different detunings. The other two are $\{\Delta_{ph}=\pm\Delta_v,\quad\Delta_{pv}=\pm\Delta_h\}$, which impose crossed constraints between different wavelengths, and are therefore unlikely to occur. The fourth solution can be written as
\begin{equation}
\Delta_{ph}=-\Delta_{pv}\;, \quad \quad \Delta_h=-\Delta_v\;.\label{typeii}
\end{equation}
Such condition is sketched in figure \ref{astigmatism}. 
When the first order transverse mode is excited in the signal polarization (extraordinary) the astigmatism is small ($\Delta_h - \Delta_v\approx 0.6$). This situation is depicted in figure \ref{astigmatism}a. 
When the first order transverse mode is excited in the idler polarization (ordinary) the astigmatism is large ($\Delta_h - \Delta_v\approx 4.3$). This situation is depicted in figure \ref{astigmatism}b. As explained in the experimental section, the astigmatic mode splitting can be tuned by varying the crystal orientation and temperature.

Two different dynamics result from these conditions. The signal polarization can sustain the simultaneous operation of $h$ and $v$ modes, and acquire OAM. We refer to this dynamical regime as OAM tunning.
However, the idler polarization cannot receive OAM because its $h$ and $v$ resonances are too far apart for being simultaneously excited. In this case, only one of the HG components can survive and the transverse mode of the idler polarization may abruptly switch between $h$ and $v$ as the crystal is tuned. We refer to this dynamical regime as HG switching. 
In the next sections, we address this point by evaluating the oscillation threshold in each case and performing numerical solutions of \eqref{eq-5OPO1}-\eqref{eq-5OPO5}. We show that our experimental results agree with the present model in two limiting cases for the detuning parameters.

\section{HG mode switching and OAM tuning}

When different operation regimes compete, the outcome of the dynamical equations is strongly influenced by the relative magnitudes of the threshold power associated with each regime. 
Thus, let us initially consider the threshold obtained from \eqref{idler-h} and \eqref{idler-v} for a LG$_{01}$ pump ($\theta=\pi/2$ and $\phi=\pi/2$). 
For each component, it is given by
\begin{eqnarray}
I^{th}_{h} &=& I^{th}_{v} = \frac{I_{res}^{th}}{2}(1+\Delta^2)(1+\Delta_p^2) ,\, \label{5threshold}
\end{eqnarray}
where $\Delta_p\equiv\Delta_{ph}=-\Delta_{pv}$ and the astigmatic mode splitting is given by $\Delta\equiv\Delta_h=-\Delta_v$ due to \eqref{typeii}. Notice, however, that there are two different regimes for the nonlinear process. The transverse components can either cooperate in a five-mode coupling when the HG resonances are close enough (figure \ref{astigmatism}a), or oscillate in independent three-mode channels when the HG resonances are too far apart (figure \ref{astigmatism}b). 
For the three-mode operation ($h-h-00$ or $v-v-00$) with resonant down-converted beams, the threshold is
\begin{equation}
I^{th}=I_{res}^{th}(1+\Delta_p^2) \,.\label{3threshold}
\end{equation}
Comparing \eqref{3threshold} and \eqref{5threshold} we identify the two possible operating regimes. If $|\Delta| < 1$, \eqref{5threshold} is always lower than \eqref{3threshold} and the five-mode operation is favorable. For greater values of $\Delta$, the three-mode operation may dominate.

\begin{figure*}
	\includegraphics[scale=0.25]{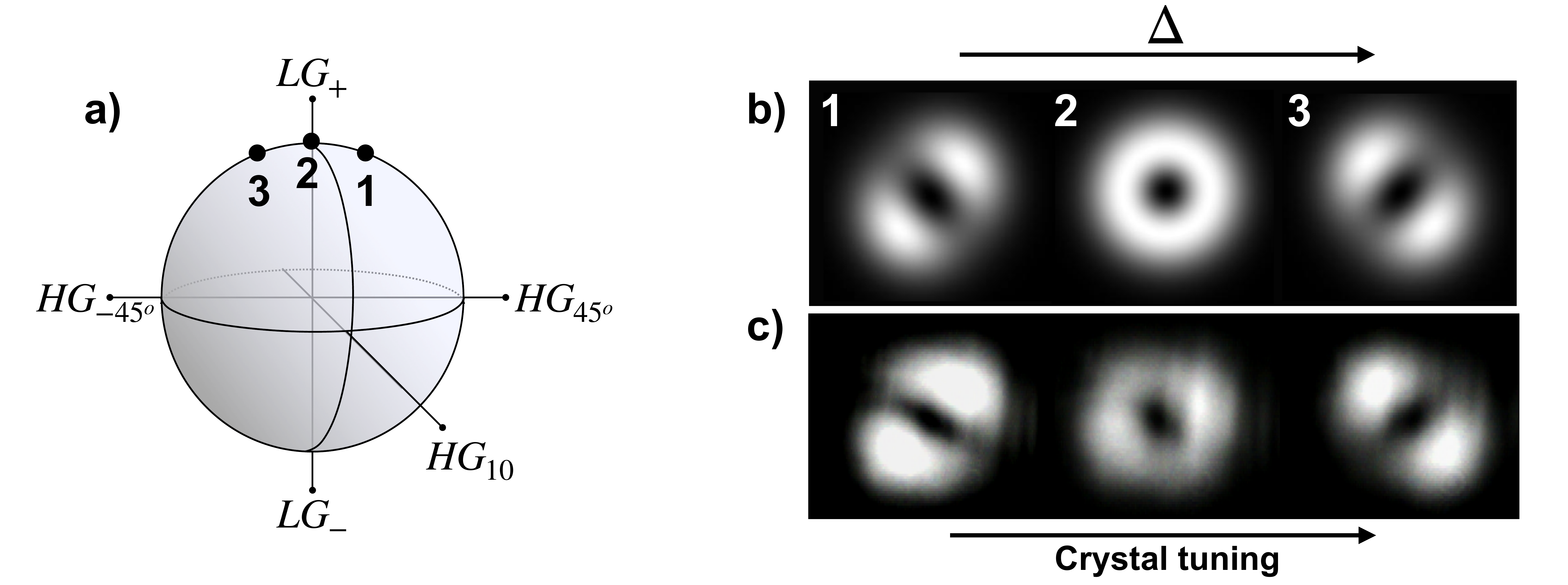}
	\caption{Experimental  and theoretical results of the OAM tuning. In (a) we show the numerical solutions for the steady-state signal mode in the Poincaré representation for 1) $\Delta=-0.5$ 2) $\Delta=0$ and 3) $\Delta=0.5$. The corresponding intensity profiles are shown in (b). The considered pump detunings are $\pm\Delta/20$ and the pump power is 1.5 times the threshold at resonance. The experimental results are shown in (c) as the crystal is smoothly tuned.}
	\label{OAMresult}
\end{figure*}
 
When the astigmatic splitting is small ($|\Delta|<1$), 
the five-mode regime takes place and opens an interesting possibility regarding fine-tuning of the OAM transfer. This corresponds to the case where the signal polarization operates in a first-order mode, while the idler is in the TEM$_{00}$ mode. In fact, 

the phase difference between the $h$ and $v$ components of the signal polarization depends on the astigmatic splitting $\Delta\,$, which can be tuned by playing with the crystal orientation and temperature.  In figures \ref{OAMresult}a and b, we show the numerical results of \eqref{eq-5OPO1}-\eqref{eq-5OPO5} for the extraordinary mode for three different values of $\Delta\,$. It can be seen that the latitude of the signal mode in the Poincaré sphere can be accurately controlled, evidencing the fine-tuning of the OAM transfer. The experimental results shown in figure \ref{OAMresult}c are in good agreement with the theoretical prediction. 

\begin{figure}[htbp]
	\includegraphics[scale=0.245]{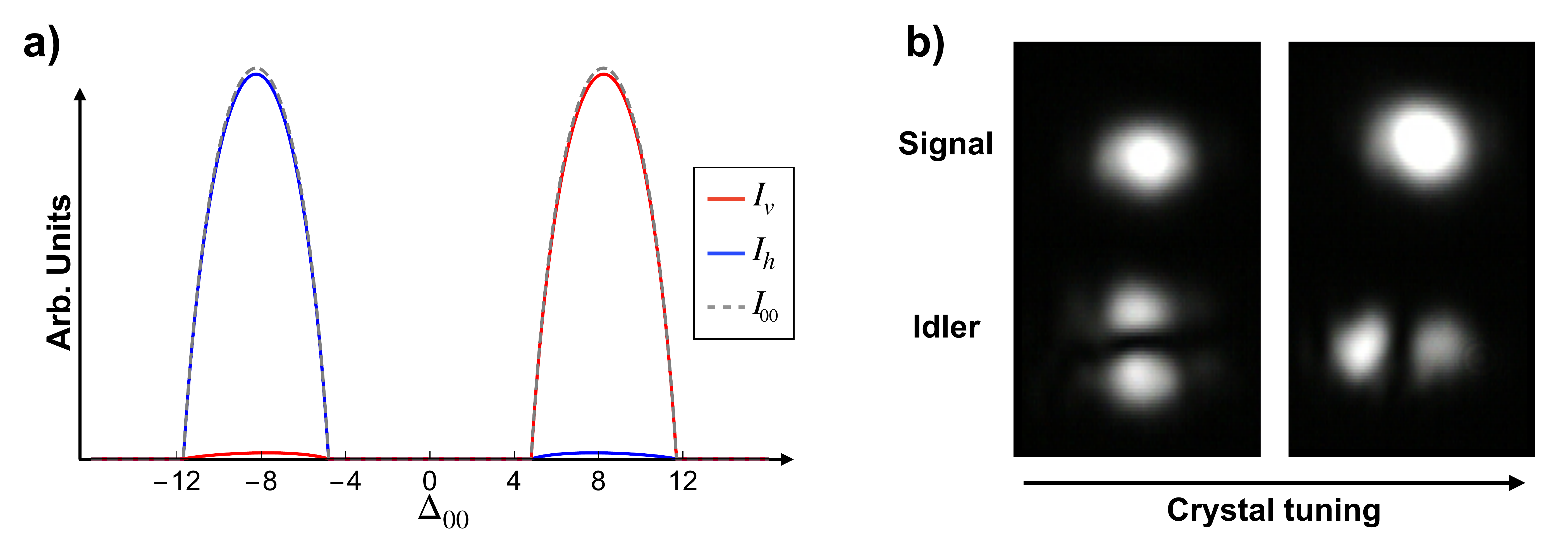}
	\caption{Experimental and theoretical results of the HG switching. In (a) we show the numerical simulations of the down-converted intensities as a function of $\Delta_{00}$ under condition \eqref{typeii}. The parameters used are $\Delta=8$, $\Delta_p=0$, $\theta=\phi=\pi/2$ and the pump input is twice the threshold at resonance. The experimental results are presented in (b), showing the sharp mode switching under crystal tuning.}
	\label{OAMvsHG}
\end{figure}

Another observed regime is the three-mode operation that takes place when the astigmatic splitting is large ($|\Delta|>1$). This is the case when the signal is a TEM$_{00}$ and the idler is a first order mode. In figure \ref{OAMvsHG}a we show the numerical solutions of the dynamical equations as a function of $\Delta_{00}\,$. One can see that there is no OPO around the OAM tuning region $\Delta_{00}=0\,$, but the three-mode operation takes place as $\Delta_{00}$ approaches 
$\pm\Delta\,$. In this case, OAM tuning is not possible and changes in the crystal parameters only lead to abrupt switching between the \textit{h} and \textit{v} channels. Two images corresponding to this HG mode switching are shown in figure \ref{OAMvsHG}b.

\section{Conclusion} \label{conclusion}
The dynamics of structured beams in an optical parametric oscillator is determined by a compromise between the spatial overlap of the interacting modes and the boundary conditions that affect their resonances. Transfer of orbital angular momentum is strongly dependent on the interplay between these parameters and can be controlled through a suitable tuning of the cavity and crystal conditions. We described two situations in which the anisotropy can be tuned either to allow orbital angular momentum transfer or to switch between different Hermite-Gaussian profiles. These achievements improve our understanding of the relevant effects behind the parametric amplification of structured light and can be useful for engineering quantum information devices based on twisted beams.

\onecolumngrid

\vspace{5mm}
\twocolumngrid

\section*{Acknowledgments}
Funding was provided by Coordena\c c\~{a}o de Aperfei\c coamento de Pessoal de N\'\i vel Superior (CAPES), Funda\c c\~{a}o Carlos Chagas Filho de Amparo \`{a} 
Pesquisa do Estado do Rio de Janeiro (FAPERJ), Instituto Nacional de Ci\^encia e Tecnologia de Informa\c c\~ao Qu\^antica (INCT-IQ) and Conselho Nacional de Desenvolvimento Cient\'{\i}fico e Tecnol\'ogico (CNPq).

\bibliographystyle{unsrt} 
\bibliography{references} 

\end{document}